\documentstyle[11pt,aaspp4]{article}

\def\chandra{{\it Chandra}}
\def\lsim{\hbox{\raise.35ex\rlap{$<$}\lower.6ex\hbox{$\sim$}\ }}

\received{29 September 1999}
\accepted{26 October 1999}

\slugcomment{Accepted by the Astrophysical Journal}

\lefthead{First Light \chandra\ Observation of Cas A}
\righthead{Hughes, Rakowski, Burrows, \& Slane}

\begin{document}

\title{Nucleosynthesis and Mixing in Cassiopeia A}

\author{
John P.~Hughes\altaffilmark{1,2}, 
Cara E.~Rakowski\altaffilmark{1,2},
David N.~Burrows\altaffilmark{3}, and
Patrick O.~Slane\altaffilmark{4}
}
\altaffiltext{1}{Department of Physics and Astronomy, Rutgers The State
 University of New Jersey, 136 Frelinghuysen Road, Piscataway NJ 08854-8019;
 E-mail: jph@physics.rutgers.edu and rakowski@physics.rutgers.edu
}

\altaffiltext{2}{
Also Service d'Astrophysique, L'Orme des Merisiers, 
CEA-Saclay, 91191 Gif-sur-Yvette Cedex France
}

\altaffiltext{3}{
Department of Astronomy \& Astrophysics, 525 Davey Lab,
Penn State University, University Park, PA 16802  
}
\altaffiltext{4}{
Harvard-Smithsonian Center for Astrophysics, 60 Garden Street,
Cambridge MA 02138  
}

\begin{abstract}

We present results from the first light observations of the Cassiopeia
A (Cas A) supernova remnant (SNR) by the \chandra\ X-ray Observatory.
The X-ray spectrum varies on all spatial scales down to the
instrumental limit (0.02 pc at the SNR). Based on representative
spectra from four selected regions we investigate the processes of
nucleosynthesis and mixing in Cas A.  We make the first unequivocal
identification of iron-rich ejecta produced by explosive
silicon-burning in a young Galactic SNR. Elsewhere in the remnant we
see silicon-rich ejecta from explosive oxygen-burning.  Remarkably,
our study finds that the Fe-rich ejecta lies outside the Si-rich
material, leading to the conclusion that bulk motions of the ejecta
were extensive and energetic enough in Cas A to cause a spatial
inversion of a significant portion of the supernova core during the
explosion.  It is likely that this inversion was caused by ``Fe''-rich
ejecta emerging in plumes from the rising bubbles in the
neutrino-driven convection layer.  In addition the radioactive decay
energy from $^{56}$Ni may have contributed to the subsequent evolution
of the material.  We have also discovered faint, well-defined
filaments with featureless X-ray spectra that are possibly the sites
of cosmic ray acceleration in Cas A.

\end{abstract}


\section{INTRODUCTION}

\par

Young supernova remnants (SNRs) are the critical link between the
nucleosynthetic processes that occur in stars and essentially all the
metals that exist in the Universe.  Accurate knowledge of the
nucleosynthetic yields from exploded stars is essential for studies of
the evolution of the interstellar medium, external galaxies, and even
clusters of galaxies. However, models of nucleosynthesis have been
tested almost exclusively in the ensemble (i.e., averaged over a
stellar initial mass function) by comparison against meteoritic and
solar photospheric abundances (e.g., Thielemann, Nomoto, \& Hashimoto
1996, hereafter TNH).  Direct comparison of the models against
abundance data derived from individual supernovae (SNe) or their
remnants has been much more limited (e.g., Hughes \& Singh 1994).

\par 

Numerical models (e.g., Woosley \& Weaver 1995; TNH) predict that
nucleosynthesis in core-collapse SN occurs in a layered,
``onion''-skin type manner.  Near the core of the star where the
burning shock temperatures are highest, explosive Si-burning
completely exhausts Si and results in a composition that is dominated
by $^{56}$Ni, which, as it decays to $^{56}$Co and $^{56}$Fe, powers
the light curve of the supernova.  Further out from the center of the
star the shock temperature has dropped and explosive Si-burning is
incomplete.  In this zone significant amounts of what will ultimately
become Fe are still produced, but a large fraction of the matter is
left in the form of Si, S, Ar, and Ca.  At lower temperatures and
further out still, explosive O-burning occurs, leading to a
composition dominated by O and Si with very little or no Fe.  Finally,
there is explosive Ne/C-burning which produces mostly O.  At this
point, the temperature of the burning shock has fallen far enough that
the composition of layers of the star above are unaltered by the SN
explosion.  Instead they reflect the nucleosynthesis that occurred
during the normal course of the star's hydrostatic evolution.

\par 

The high sensitivity, broad bandwidth, good spectral resolution, and
unsurpassed angular resolution of the recently launched \chandra\
X-ray Observatory (see Weisskopf et~al.~1996 for a description) has
opened up a new window on the explosive nucleosynthesis process in
supernovae. It is now possible to measure the composition of
individual knots within SNRs and, by comparison to models of the
nucleosynthetic process, determine in what layer or layers of the
progenitor star the material was formed. In this letter we report on
early \chandra\ observations of the Cassiopeia A (Cas A) SNR in which we
demonstrate for the first time this powerful new technique.

\par

Cas A is widely believed to be the result of a core collapse supernova
explosion of a massive star (Fesen et~al.~1987) that was possibly
witnessed by Flamsteed in 1680 and was rediscovered as a remnant in the
radio by Ryle \& Smith (1948) over 250 years later.  At a distance of
3.4 kpc (Reed et~al.~1995), the $2^\prime$ radius optical shell
corresponds to a physical size of 1.7 pc. In the optical band, Cas A
displays complex variations of composition with position and velocity,
and it is one of the brightest SNRs at X-ray and radio
frequencies.

\section{OBSERVATIONS AND DATA REDUCTION}

Cas A was observed for 6,100 s by ACIS-S (Garmire 1997) on 20 August,
1999, during Orbital Activation and Check-out of \chandra.  A
processed event file of the observation (with creation date
1999-09-04T03:08:28) was obtained from the \chandra\ X-ray Center and
analysis was carried out using standard astronomical software. The
spectra were extracted using the energy column from the \chandra\
event file and binned into energy channels containing at least 25
events each.
At this point the calibration of the instrument is still preliminary;
for example, the energy scale of the spectral data may be uncertain by
tens of percent. Nevertheless the bright, well-known spectral features
in the remnant's spectrum allow for the nearly unambiguous
identification of emission from the astrophysically abundant elements
Mg, Si, S, Ar, Ca, and Fe. For comparison to observed spectra we use
preliminary response functions for the effective area of the telescope
(acis\_bi.arf) and the efficiency and spectral resolution of the
ACIS-S instrument (w134c4r\_norm.rmf) convolved with simple models for
the X-ray emission from SNRs, including the effects of nonequilibrium,
or time dependent, ionization (NEI) (Hughes \& Singh 1994).  In this
letter we aim for a qualitative description of several representative
\chandra\ spectra; quantitative studies with formal $\chi^2$-type fits
to the data are deferred to future work.

\section {DATA ANALYSIS AND DISCUSSION}
\subsection{Image Analysis}

A color image of Cas A, encoding information on the intrinsic X-ray
spectrum of the remnant, is shown in Figure 1.  Regions that appear
red indicate places where Cas A shows more emission from the energy
band (0.6--1.65 keV) that contains the K-shell lines (transitions to
the 1s electronic level) of O, Ne, and Mg and L-shell lines
(transitions to the 2s electronic level) of Ca and Fe.  Green regions are
relatively enhanced in the K-shell emission lines of Si (energy band
1.65--2.25 keV), while blue regions are weighted toward higher energy
emission (2.25--7.50 keV) that includes the S, Ar, Ca, and Fe K-shell
emission lines.  Regions with comparable levels of emission in the
three bands appear white.  All three bands also contain emission from
thermal bremsstrahlung as well as other nonthermal continuum
components.

\par

The overall, gross morphology of Cas A has been known from earlier
studies, but the exquisitely fine spatial details of its X-ray
emission apparent in Fig.~1 (and Fig.~2, the broadband total intensity
X-ray image) are entirely without precedent.  Furthermore, the
\chandra\ data show that the X-ray spectral character of the SNR
varies on all angular scales down to the resolution limit of the
telescope ($\sim$1$^{\prime\prime}$), which corresponds to physical
scales of $\sim$0.02 pc at the remnant. Features of the X-ray remnant
to note (colors refer to Fig.~1) are (1) the outer blast-wave, which
fully surrounds the remnant and appears reddish-purple; (2) diffuse
clumps of reddish emission most obvious to the east and north; (3)
bright compact knots (largely green or white-colored) within the
bright shell; and (4) diffuse filamentary emission pervading the
interior, which varies considerably in color.
As we illustrate below, using spectra from four representative
regions, spectral differences arise from variations in the underlying
radiation process, composition, excitation conditions, and the column
density of absorbing gas and dust in the line-of-sight toward Cas A.

\subsection{Spectral Analysis of Stellar Ejecta}

The compact, high surface brightness knots are among the most striking
new features of the \chandra\ Cas A image.  In Figure 3 we show the
spectrum of one such feature, namely the bright green-colored knot
toward the SE labeled A in Fig.~2.  The dominant spectral feature is
the (unresolved) complex of K-shell transitions from the helium-like
ion of Si at a photon energy of $\sim$1.85 keV although the spectrum
also contains emission from other elemental species: Mg, S, Ar and Ca.
This knot displays the largest equivalent width Si line we have yet
detected in the remnant, $W_E = 1.6\,\rm keV$, more than twice the
maximum equivalent width expected from a hot plasma with solar
abundances under NEI conditions, $W_E = 0.6\,\rm keV$.  On the other
hand, the equivalent width of the Si line is far smaller than the
value expected from a knot of pure Si or even a knot of pure metal
ejecta composed only of those elemental species with observed K-shell
line emission.  A number of other compact features in the remnant
share these characteristics, although with slightly less dominant Si
lines, including many of the white unresolved knots in the northern
section of the bright shell.

\par

The enhanced Si and S abundances, relative to solar, identify these
compact knots as stellar ejecta from deep within the progenitor star
where explosive O-burning or incomplete Si-burning took place (TNH).
The elemental species with K-shell lines in the observed band are
unable to produce sufficient continuum emission to explain the amount
seen and so another source for the continuum is required.  Additional
material with low enough atomic number ($Z<8$) that its emission lines
fall below the observed band is required (a solar abundance plasma
produces too much Fe L-shell emission). The spectral model plotted
against the spectrum from region A as the solid line in Fig.~3 is a
representation of this situation: it contains the products of
explosive O-burning in the predicted ratios from TNH (taken from their
Table 2) plus a H/He continuum (a N-rich plasma would work as well for
the continuum).  Parameters describing the thermodynamic state of the
plasma (temperature, ionization timescale, and line-of-sight absorbing
column density) were adjusted for broad agreement with the observed
line shapes.  All of the obvious line features in the spectrum from
this region can be described by emission from species that are
predicted to be produced in explosive O-burning.  In particular we
note the low energy features from Ca L-shell lines (0.65 keV and 0.85
keV) and Mg K-shell emission (1.34 keV).  Both of these species are
observed to be somewhat more abundant than predicted by TNH.  On the
other hand, emission from Fe is well constrained by the lack of both
L- and K-shell emission (dotted curve), which argues strongly against
an origin in the incomplete Si-burning zone.  We conclude therefore
that the bright, compact knots are stellar ejecta dominated by the
products of explosive O-burning that have been partially mixed with
lower-$Z$ material from layers of the star further out in radius.

\par

Until now it has not been possible to obtain a comprehensive picture
of nucleosynthesis in Cas A because of the lack of observed emission
from newly synthesized iron in the ejecta. Fe is produced in the
deepest layers of the star during explosive Si burning, which,
depending on the temperature of the burning front, can range from
complete (resulting in nearly pure Fe) to incomplete (resulting in a
spectrum of nucleosynthetic products from Si to Fe).  In order to
address this issue we examined two regions in Cas A that show evidence
(Vink et~al. 1999) for Fe K-shell emission at $\sim$6.7 keV and are
spectrally different from each other and from the compact Si- and
S-rich knots.  The first, region B, is on the western edge of the
remnant and appears strikingly blue in Fig.~1, while the second
(C) covers a faint knot on the eastern side (red color in
Fig.~1).

\par

Previous work (e.g., Holt et~al.~1994; Keohane et~al.~1996; Vink
et~al.~1999) has noted the hardness of the X-ray emission from region
B and the \chandra\ data (Fig.~3) appear to support this.  The
spectrum is thermal; compared to region A it is more highly absorbed
and the ionization state is more advanced (i.e., note the higher
intensity of the Si hydrogen-like Lyman$\alpha$ line at 2.0 keV
compared to the helium-like lines at $\sim$1.85 keV). The
thermodynamic state of the plasma was set to match the Fe L-shell
complex at 1--1.5 keV, the Fe K-shell line at $\sim$6.7 keV, and the
line shapes of the other bright lines.  Given the state of
instrumental calibration and the simplicity of our fits, the agreement
between data and model is excellent.  The most remarkable feature of
the comparison is the inferred abundance set. The model plotted in
Fig.~3 contains only the products of incomplete explosive Si-burning
in the predicted ratios from TNH, plus continuum from lower
$Z$-material. To repeat: the relative abundances of Si, S, Ar, Ca and
Fe were not fitted parameters.

\par

On the other side of the remnant is region C whose spectrum is
dominated by a broad, emission feature around 1 keV.  At higher
energies there is a rapidly falling continuum, modest lines from Si
and S, and an Fe K-shell line with an equivalent width of
$W_E\sim2000$ eV.  This is a bizarre spectrum. The only abundant
chemical species that can produce quasi-continuum emission around 1
keV is Fe. Our model simulations, using relative abundances from
incomplete Si-burning, confirm this (dotted curve). However these fits
also show that the incomplete Si-burning abundance set grossly
overpredicts the Si, S, Ar, and Ca emission.  A much improved fit
(solid curve) results when the abundances of these species are reduced
by factors of 5 or more, indicating that the composition of the knot
is considerably more Fe-rich and therefore closer to that predicted
for complete Si-burning. The Fe K-shell emission, as well as the
shapes of the other K-shell lines, are fit quite well by this
model. Differences near 1.5 keV between the data and simulations are
due to the well-known deficiencies of plasma model codes
for predicting the precise shape of the Fe L-shell emission
(Brickhouse et~al.~1995).  The overall flux, which can be more
accurately modeled, is in good agreement.  As for both cases before,
this spectrum also requires an additional source of continuum
emission.

\subsection{Mixing and Bulk Motion of Ejecta in Cas A}

Amazingly, according to the inferred composition, the X-ray data are
probing the explosive nucleosynthetic processes that occur at the
highest temperatures in an supernova, i.e., the Fe- and Si-rich
ejecta. Perhaps even more remarkable is the difference in the expected
locations of the products of O-burning (Si) and Si-burning (Fe).  In
Cas A, the most Fe-rich material lies at the outermost edge of the
ejecta, while the Si-rich ejecta is generally closer to the center.
This is best seen on the eastern side of the remnant, where fingers of
Fe-rich material (red color in Fig.~1) clearly precede the Si-rich
knots (whitish color in Fig.~1).  (The lack of Si beyond the Fe
emission, as well as a lack of Fe inside the Si-emission, argues
against projection effects.)  The conclusion is obvious: the ejecta in
Cas A have undergone a spatial inversion or overturning of the
explosive O- and Si-burning products. It is also clear, however, that
this process did not lead to a homogenization of this portion of the
ejecta, since we see individual knots produced by the different
burning processes.

\par

Previous evidence (Fesen \& Gunderson 1996) for mixing of inner and
outer layers of Cas A has come from optical spectroscopy of knots in
the ``jet'' toward the northeast, which can be seen in the \chandra\
image as a few faint linear wisps of X-ray emission.  The presence in
a dozen or so optical knots of explosive O-burning products (O, S, and
Ar) that also contained H and N from the star's outer layers was taken
as a sign that mixing had occurred between the O-burning zone and the
hydrostatic layers of the star.  We appear to require this type of
mixing as well, since all three of our extracted spectra require
additional low $Z$-material to explain the observed level of continuum
emission (note that some fraction of the continuum may be coming from
other processes, see \S3.4).  However, our results also require the
bulk motion of significant fractions of the Si-burning zone out past
the O-burning zone during the explosion that formed Cas A.  This
inversion of the nucleosynthetic products points to a considerably
more extensive, energetic, and violent process of large-scale mixing
than previously observed, although mixing of this type has been
proposed to explain the light curve and the early appearance of X-ray
and $\gamma$-rays from SN1987A (see Woosley 1988 and references
therein).

\par

The theoretical basis for large scale mixing and bulk motions of
ejecta has also been established recently.  Modelers now recognize the
importance of neutrino-driven convection for initiating core-collapse
SN explosions (Herant et~al.~1994; Burrows, Hayes, \& Fryxell 1995).
Multidimensional radiation/hydrodynamics simulations of this process
show that it results in a massively aspherical explosion that includes
``crooked fingers'' of ejecta blown out at high speeds.  Homogenization
of the material is not expected because any single parcel of core
matter undergoes neutrino-driven convection for only one or two cycles
of rising and falling.  At later times, the radioactive decay energy
of $^{56}$Ni ($\sim$$10^{17}$~ergs~gm$^{-1}$), which is comparable to
the kinetic energy density in the ``Fe''-rich ejecta
($\sim$$2\times10^{16}$~ergs~gm$^{-1}$) (Woosley 1988), may play a
role.  We note that the bright compact knots in the remnant are nearly
all Si-rich (the bright whitish and greenish structures in Fig.~1),
while the knots of Fe-rich ejecta (red features) tend to be more
diffuse and extended.  This is suggestive of an additional heating
source in the Fe-rich knots which could be the radioactive decay
energy.  However, until theoretical studies tell us how the formation,
evolution, and motion of ejecta knots depends on their position in the
star, it will be difficult to quantify the amount of additional
heating required to explain the more diffuse nature of the Fe-rich
knots.

\subsection{Featureless Continuum Emission}

Spectra of enriched stellar ejecta alone do not provide a complete
picture of Cas A. Some regions within the remnant show no evidence for
thermal emission; their spectra are devoid of line emission.  Region D
contains one such filament that we selected based on its curious
purple color suggesting an intrinsic spectrum weak in Si K-shell
emission.  Indeed the extracted spectrum is nearly featureless; the
equivalent widths of the Ne, Mg, Si, S, Ar, Ca, and Fe K-shell lines
are all less than 200 eV. It is well described by an absorbed
power-law model with a photon index of $\alpha \sim 2.6$, but could be
fit equally well by an absorbed thermal continuum with $kT \sim 2.5$
keV. For the filament to be purely thermal it would need to have very
low abundances of the species from Ne through Fe. One possibility is
that we are seeing a portion of the blast wave in the ISM in which the
metals have been depleted onto dust grains (e.g., Vancura et
al.~1994).  This is unlikely, since some chemical species, Ne and Ar
especially, are not expected to be depleted. Furthermore, the
\chandra\ spectrum of the blast wave (the faint outer shelf of
emission) does clearly show significant Si and S emission lines.
Another possibility is that this filament is thermal continuum from
low $Z$ material, similar to what we argue above underlies the
thermal, metal-rich spectra of the several ejecta knots that we
studied.

\par

It is more likely that this filament is emitting X-ray synchrotron
radiation from high energy electrons that have been accelerated to TeV
energies in the SN shock like in SN1006 (Koyama et~al.~1995) and
G347.3$-$0.5 (Koyama et~al.~1997; Slane et~al.~1999).  The photon
index is comparable to the values measured from these other SNRs
($\sim$1.9 and $\sim$2.5, respectively) in which nonthermal
synchrotron radiation dominates the X-ray emission in the 2-10 keV
band.  Furthermore, Allen et~al.~(1997) inferred the presence of
nonthermal power-law emission from Cas A underlying the predominantly
thermal emission in the band below 16 keV.  Their power-law index,
$\alpha=1.8^{+0.5}_{-0.6}$, is broadly consistent with what we find
here, especially given the simplicity of their model for the thermal
emission and our calibration uncertainties.  This raises the
possibility that some or even all of the continuum emission we 
argued for above is nonthermal, rather than thermal.

\par

\section{SUMMARY}

We find that the Cas A SNR contains compact, high surface brightness
knots enriched in Si and S that are consistent with the
nucleosynthetic products of explosive O-burning. Moreover the observed
equivalent widths of the line features in the knots require the
presence of continuum emission from material with low atomic number
($Z<8$) indicating that the core and mantle of the star were mixed
during the SN explosion, as also found by others.  Elsewhere in the
remnant we find more diffuse, lower surface brightness features with
abundances that are consistent with explosive Si-burning. These
features contain a considerable amount of Fe, which in some cases
dominates the X-ray emission, and this is what differentiates them
from the O-burning products.  Remarkably these Fe-rich features lie at
the outer edge of the bright regions that mark the distribution of the
SN ejecta. Since explosive Si-burning occurs at greater depths in the
progenitor star than O-burning does, these results require that a
spatial inversion of a significant portion of the star's core has
occurred.  This was likely the result of neutrino-driven convection
during the initiation of the SN explosion.  The radioactive decay
energy of $^{56}$Ni may explain the more diffuse nature of the Fe-rich
ejecta knots.  Finally we have discovered faint, sharply defined
filaments with nearly featureless spectra that are likely to be
emitting nonthermal radiation and may be indicating the sites of
cosmic ray shock acceleration in Cas A.  This brief report barely
touches on the wealth of information and new discoveries that will be
forthcoming as we study the \chandra\ data on this and other SNRs.

\acknowledgements

This work would not have been possible without the dedication,
sacrifice, and hard work of the literally thousands of scientists,
engineers, technicians, programmers, analysts, and managers who worked
on the \chandra\ X-ray Observatory during the long course of its
development.  Special thanks are offered to them. Helpful discussions
with Anne Decourchelle and Thomas Douvion on the scientific content of
the article are gratefully acknowledged.  We appreciate Monique
Arnaud's support and hospitality during the course of this project.
This work was partially supported by NASA Grant NAG5-6420.

\vfill\eject

\newpage

\clearpage

\figcaption[f1.ps]{\chandra\ X-ray image of Cas A, encoding
information on the intrinsic X-ray spectrum. The figure was
constructed from images in three different energy bands: 0.6--1.65 keV
(red), 1.65--2.25 keV (green), and 2.25--7.50 keV (blue). The images
contained a total of $4.1\times10^5$, $2.9\times10^5$, and
$1.9\times10^5$ X-ray events with peak number of events (in any single
$1^{\prime\prime}$ square pixel) of 59, 89, and 47, respectively.  The
separate images were square-root scaled from 0 to 80\% of their
individual peak values after being smoothed to a signal-to-noise ratio
per pixel of 7 by a Gaussian function whose width depended on the
local image intensity. The main effect of the smoothing process was to
make the faint outer regions of the remnant visible; the bright,
high spatial frequency parts of the images were unaffected.
\label{Figure 1}}

\figcaption[f2.ps]{Broadband \chandra\ X-ray image of Cas A. The
spectral extraction regions in our study are indicated.  
\label{Figure 2}}

\figcaption[f3.ps]{Energy spectra from several regions in Cas A as
indicated in figure 2.  The horizontal error bars show the widths of
the energy bins and the vertical ones indicate the statistical error
on the measured event rate; systematic errors are not included.
Superposed on the data points are smooth curves of simulated \chandra\
ACIS-S spectra.  The simulations for regions A, B and C are of a
shock-heated plasma with nonequilibrium ionization fractions absorbed
by line-of-sight interstellar material.  The dotted curves in regions
A and C, and the solid curve for region B, assume abundances
corresponding to explosive incomplete Si-burning. A considerably
better match for region A uses O-burning abundances (solid curve).
The solid curve for region C is more Fe-rich, the Si, S, Ar, and Ca
abundances are reduced by factors of 5 or more from their values in
incomplete Si-burning.  The solid curves for regions A, B, and C have
temperatures of 2.5 keV, 2.5 keV, and 2.8 keV, ionization timescales
of $2.5\times10^{10}$~cm$^{-3}$~s$^{-1}$,
$7.9\times10^{10}$~cm$^{-3}$~s$^{-1}$, and
$7.9\times10^{10}$~cm$^{-3}$~s$^{-1}$, and column densities of
$0.9\times10^{22}$~atoms~cm$^{-2}$,
$2.3\times10^{22}$~atoms~cm$^{-2}$, and
$1.5\times10^{22}$~atoms~cm$^{-2}$ respectively.  All the models for
regions A, B and C also include significant amounts of continuum
emission from material with lower atomic number.  The solid curve for
region D is an absorbed power-law model with a photon index of 2.6 and
column density $1.3\times10^{22}$~atoms~cm$^{-2}$.
\label{Figure 3}}

\clearpage

\begin{figure}
\plotfiddle{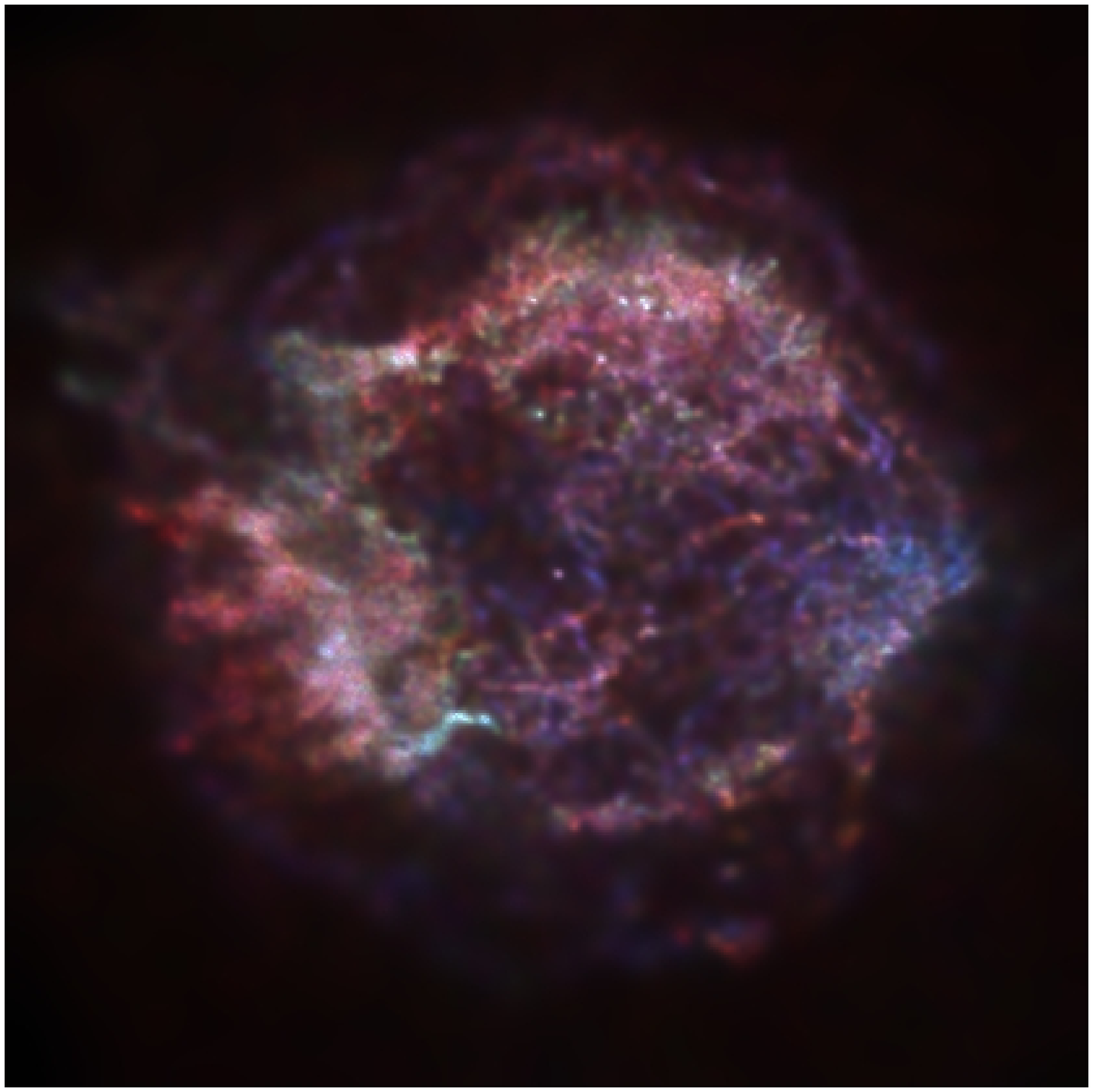}{5in}{0}{100}{100}{-300}{-200}
\end{figure}

\clearpage

\begin{figure}
\plotfiddle{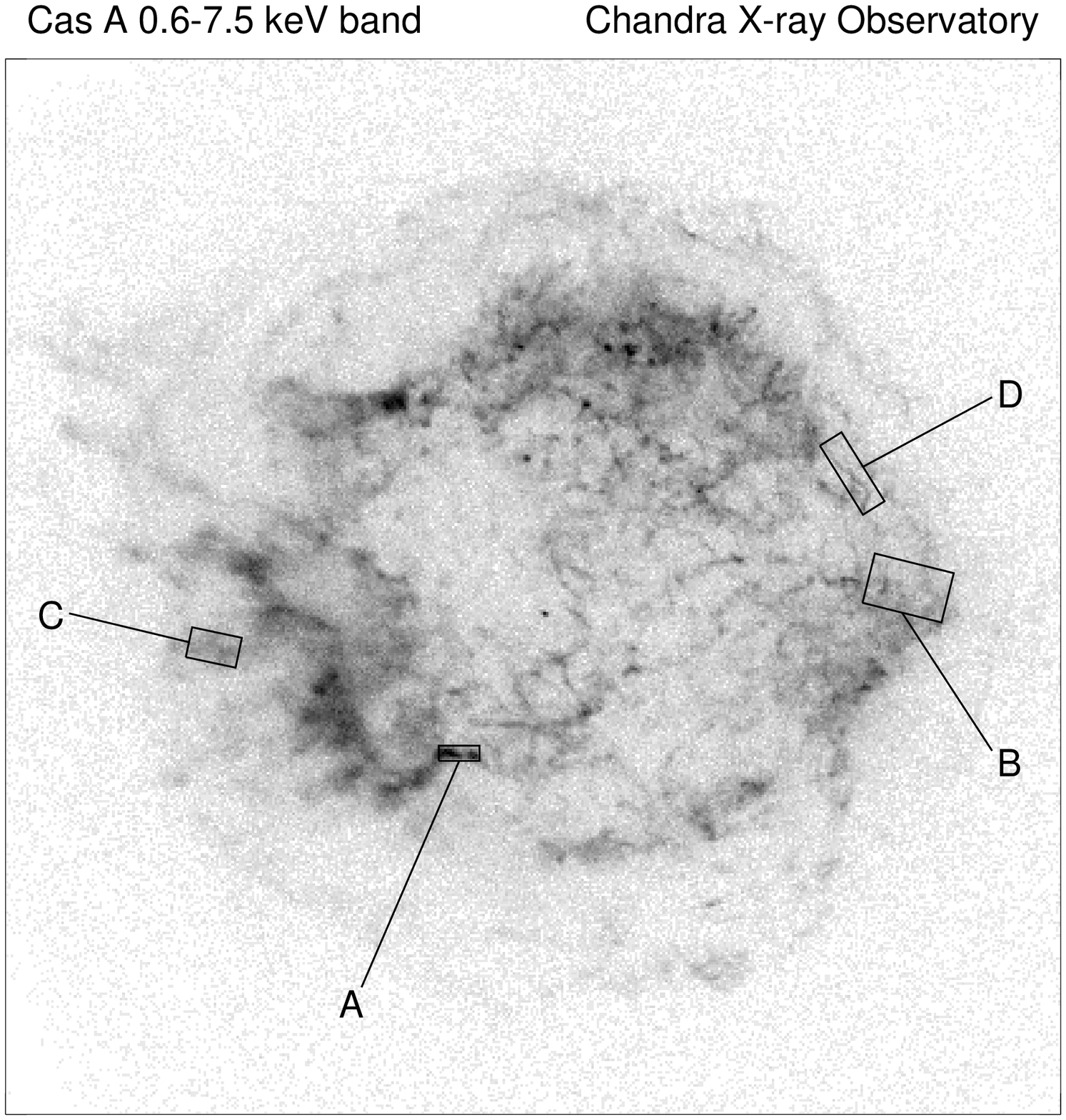}{5in}{0}{100}{100}{-300}{-200}
\end{figure}

\clearpage

\begin{figure}
\plotfiddle{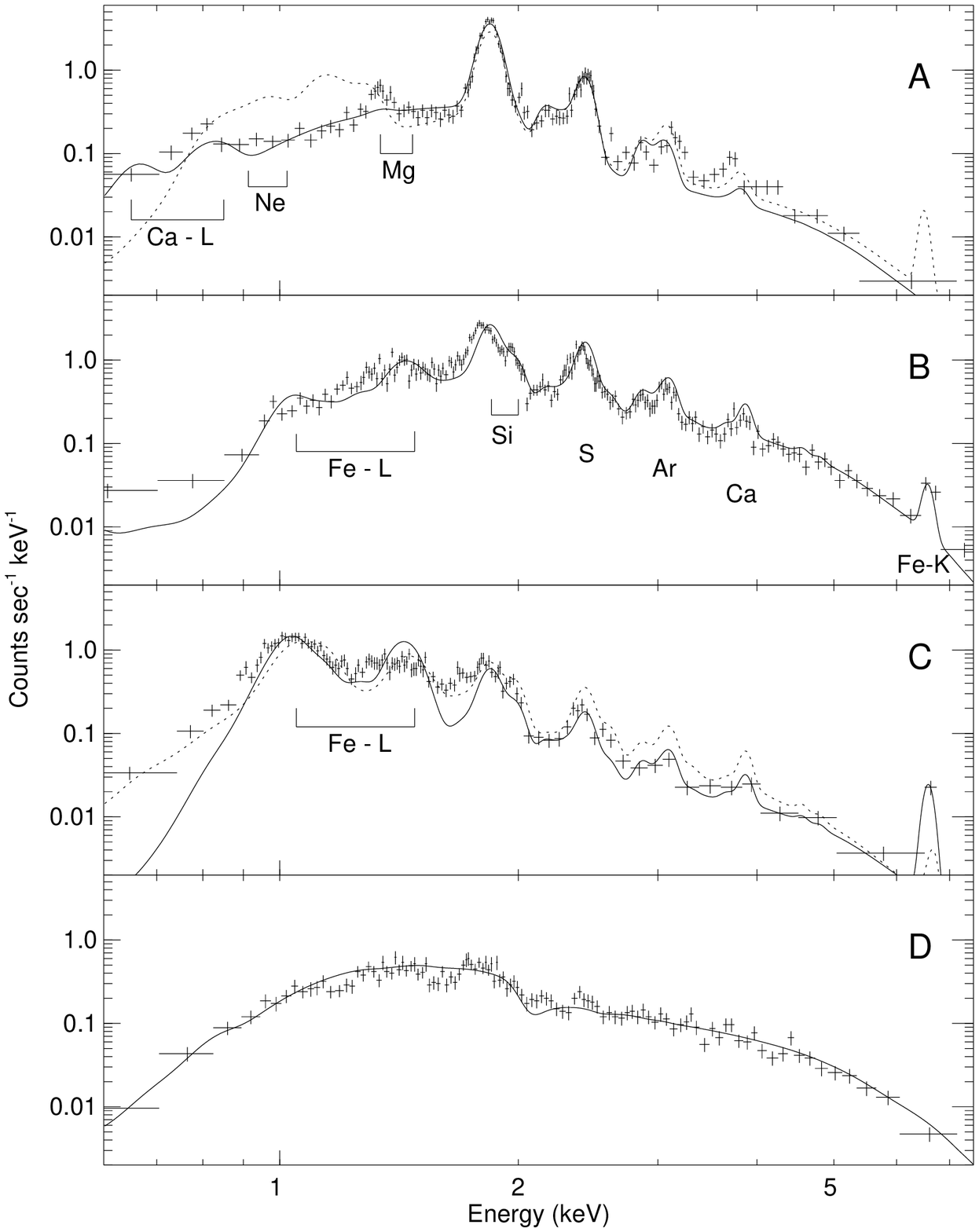}{5in}{0}{100}{100}{-300}{-250}
\end{figure}

\end{document}